\documentclass[12pt, a4paper]{article}
\usepackage[margin=1in]{geometry}
\usepackage{amsfonts}
\usepackage{amsmath}
\usepackage{amssymb}
\usepackage{bm}
\usepackage{tabu}
\usepackage[utf8x]{inputenc}
\usepackage{enumerate}
\usepackage{tgcursor}
\usepackage{xcolor, colortbl}
\usepackage{multirow}
\usepackage{graphicx}
\usepackage{placeins}
\usepackage{listings}
\usepackage{stackengine}
\usepackage{caption}
\usepackage{subcaption}
\usepackage{setspace}
\usepackage{footnote}
\usepackage{comment}
\usepackage{footmisc}
\usepackage{hyperref}
\usepackage{wrapfig}
\numberwithin{equation}{section}
\usepackage{verbatim}
\usepackage{mathrsfs}
\usepackage{natbib}
\usepackage{float}
\usepackage{algorithm}
\usepackage{algpseudocode}
\usepackage{xcolor}

\usepackage{siunitx}
\usepackage{array}
\usepackage{mathtools}

\newcommand{\vH}{\mathbf{H}}

\newcommand{\vI}{\mathbf{I}}

\newcommand{\vm}{\mathbf{m}}
\newcommand{\vM}{\mathbf{M}}

\newcommand{\vQ}{\mathbf{Q}}

\newcommand{\vR}{\mathbf{R}}
\newcommand{\vs}{\mathbf{s}}
\newcommand{\vS}{\mathbf{S}}

\newcommand{\vX}{\mathbf{X}}

\newcommand{\vY}{\mathbf{Y}}

\newcommand{\vZ}{\mathbf{Z}}

\newcommand{\vbe}{\boldsymbol{\beta}}

\newcommand{\vep}{\boldsymbol{\epsilon}}

\newcommand{\vmu}{\boldsymbol{\mu}}

\newcommand{\vnu}{\boldsymbol{\nu}}

\newcommand{\vSig}{\boldsymbol{\Sigma}}
\newcommand{\veta}{\boldsymbol{\eta}}

\newcommand{\vzero}{\boldsymbol{0}}

\makeatletter

\title{Data-Driven Modeling of Wildfire Spread with Stochastic Cellular Automata and Latent Spatio-Temporal Dynamics}

\author{ Nicholas Grieshop\footnote{\textit{University of Missouri: njgrieshop@mail.missouri.edu}}, Christopher K. Wikle \footnote{\textit{University of Missouri}}}

\begin{document}

\maketitle

\begin{abstract}
 We propose a Bayesian stochastic cellular automata modeling approach to model the spread of wildfires with uncertainty quantification.
 The model considers a dynamic neighborhood structure that allows  neighbor states to inform transition probabilities in a multistate categorical model.
 Additional spatial information is captured by the use of a temporally evolving latent spatio-temporal dynamic process linked to the original spatial domain by spatial basis functions.
 The Bayesian construction allows for uncertainty quantification associated with each of the predicted fire states.
 The approach is applied to a heavily instrumented controlled burn.
\end{abstract}


\section{Introduction}\label{sec::intro}

In 2021 there were 59,000 wildfires recorded in the USA, covering an area of over 7 million acres \citep{hoover2022wildfire}.
Anthropogenic climate change is lengthening the wildfire season and leading to an increase in wildfire frequency.
The damage caused by these fires and the expense of mitigation strategies costs billions of dollars annually.
Improving models that can predict the spread of wildfires is important for developing and managing fire fighting resources and in being able to provide timely warning to those who may be in danger from an advancing fire.

There are numerous approaches to modeling the spread of wildfires \citep[see the overview in][]{Sullivan2007AModels}. 
For example, in semi-empirical physical models such as \cite{RothermelRAFuels.}, the rate of fire spread is parameterized as a function of the heat flux into a given area and the amount of energy necessary to cause combustion.
The heat flux into a given area is a function of the fuel source and local factors such as wind speed and elevation.
These types of thermodynamic models are incorporated into wildfire simulators such as FARSITE \citep{finney1998farsite} where modifications to the approach of Rothermel are implemented, including consideration of different treatments of crown versus surface level fires and the addition of a spotting component. 
Spotting occurs when a wildfire jumps and spreads to a nonadjacent area.
Although there has been discussion of how fire spotting can be modeled from physical principles \citep[e.g.,][]{Koo2010FirebrandsFires}, incorporating realistic spotting in wildfire models remains an important topic of research.

From a mathematical perspective, wildfire propagation can be modeled using the so-called ``level set approach.''
In this method, the fire front is the object of interest, and the propagation of this front over space and time is modeled by considering the movement dynamics via an implicit level-set function. 
Information such as wind speed and elevation, as motivated by a physical model, can be incorporated into the front spread, as done in \cite{Mallet2009ModelingMethods} or \cite{Alessandri2020ParameterMethods}, where the parameters involved in the rate of spread are estimated.
In \cite{Munoz-Esparza2018AnMethod}, complex fuel sources are considered in the level set approach.
Historically, level-set methods rely on these types of empirical parameterizations and do not directly use data from the current fire to inform parameters. 
A recent exception is the work of \cite{dabrowski2022towards}, who assimilate data in real-time into a level-set model forced by a Rothermel relationship using a Bayesian filtering approach. In addition, \cite{YooLevelSet} demonstrate a hierarchical Bayesian data-driven approach to learn the spatially-varying propagation speed in the direction normal to the fire front in a level-set framework.

An alternative approach for fire spread modeling is based on cellular automata (CA), where the spread is controlled by simple rules.
In a CA, the temporal domain consists of equally spaced time points, and the spatial domain is divided into discrete cells.
CA models have a long history in applied mathematics and computer science, at least going back to \cite{JohnVonNeumannTheoryAutomata}.
In the traditional CA approach, the evolution of the states of each spatial cell from time $t-1$ to time $t$ is governed by a set of simple rules.
A famous deterministic demonstration is given by Conway's Game of Life, \cite{conway1970}, where at each time point, the transition rule for a cell considers its eight neighbors, and a state transition occurs depending on the sum of its neighbors' states.
From this simple set of rules, complex time evolution behavior can be observed without additional input from the users - aside from setting the initial state of the cells.
Two challenges in developing of a CA model are to specify the number of neighbors of a cell at time $t-1$ and to discover or specify a set of rules to update the state of the cell for time $t$.
In addition, one must decide if the transition rules are best represented as deterministic or stochastic.

Specific representations of the CA method have been incorporated into a variety of statistical models. 
For example, agent-based models in which individual agents are considered ``cells'' and operate according to a specific set of probabilistic rules have been used to model the spread of rabies in raccoon populations \citep{hooten2010statistical, wikle2015hierarchical}.
\cite{Simmonds2020TheReview} present a review of this approach applied to hydrological data and \cite{Banks2021StatisticalModeling} present a review with discussion of formal uncertainty quantification for such models.

The traditional CA approach has been applied to wildfire modeling with differing definitions of the neighborhood structure and transition rules.
For example, a traditional Moore's neighborhood (cardinal and ordinal neighbors) has often been used \citep[e.g.,][]{Karafyllidis1997AAutomata, Banks2021StatisticalModeling}.
Alternatively, \cite{Zhang2021AModels} considered a two-tiered neighborhood structure where local information was captured using a small-scale neighborhood, and an additional larger-scale neighborhood was used for long-range effects.
The spatial cells need not be defined as regular squares, as shown in \cite{HernandezEncinas2007ModellingAutomata} and \cite{Johnston2008EfficientGrid}.
This approach is further discussed in \cite{Liu2018SpreadSimulation}.

Once the neighborhood structure has been defined, the rules of state propagation must be specified or learned.
In many models, such as those discussed in \cite{Currie2019Pixel-levelSpread}, \cite{Liu2018SpreadSimulation}, \cite{Karafyllidis1997AAutomata}, and \cite{Lautenberger2013WildlandCalibration}, the rules of fire propagation are specified as simple physical relationships and are not learned.

The method outlined in this manuscript builds upon the work of \cite{Currie2019Pixel-levelSpread} by utilizing physical principles from semi-empirical models and environmental covariates when constructing a neighborhood structure. 
Importantly, rather than specify these relationships directly, our approach is data-driven in that both the neighborhood and transition rules are learned.  
In particular, we incorporate a novel dynamic neighborhood structure motivated by the \cite{RothermelRAFuels.} relationship, in addition to transition rules informed by a latent low-rank spatio-temporal dynamic process.
The model is implemented in a spatio-temporal Bayesian hierarchical framework \citep[e.g.,][]{cressie2015statistics} to provide formal uncertainty quantification of the predicted fire spread and covariate effects, analogous to the agent-based methods discussed in \cite{hooten2010statistical} and reviewed in  \cite{Banks2021StatisticalModeling}.

Section \ref{sec:propMethod} presents our approach using the CA model with physically motivated covariates and a stochastic latent process, which is then implemented in a Bayesian hierarchical model to reflect the uncertainty of data, model, and parameters. 
Section \ref{sec:Sim} gives the results of a simulation, demonstrating the CA method's ability to learn the rules of cellular evolution, and in Section \ref{sec:S5} the method is applied to a real-world fire.
Section \ref{sec:Conc} discusses further extensions to this framework and other possibilities for determining neighborhood structure.

\section{Bayesian Cellular Automata Model Methodology} \label{sec:propMethod}
This section presents the Bayesian hierarchical cellular automata model for fire spread as well as a description of the model implementation and model evaluation metrics.

\subsection{Stochastic Cellular Automata Model}\label{sec::Method-model}
Consider a fire that can occur within a rectangular spatial domain, $D_s$, that is partitioned with $n$ spatial locations (cells), $\{\vs_i: i=1,\ldots,n\}$. 
We assume the fire evolves in discrete time indexed by $t \in \{1, 2, 3, \ldots, T\}$,
and that each cell can take one of $J$ states; here we restrict our analysis to the case where $J=3$ as described below, corresponding to unburned, burning, and burned wildfire states.
We denote the state of cell $\vs_i$ at time $t$ by $\{S_t(\vs_i) = j: j=1,\ldots,J\}$.
We then assume that these potential states at each discrete time and space location follow a multinomial distribution with probabilities $p_{jt}(\vs_i)$.
That is, for each spatial location $\vs_i$ and time $t$, the state probabilities must satisfy $\sum_{j=1}^J p_{jt}(\vs_i) = 1$.
The challenge is then to estimate these spatio-temporal probabilities.

Following the method of ordered categorical responses in \cite{Albert1993BayesianData}, we relate the observed discrete ordered cateogries, $S_t(\vs_i) = j: j=1,\ldots,J$, to an unobserved latent spatio-temporal process, $Z_t(\vs_i)$. 
 Specifically, we relate the probability of an observation being in state $j$, $p_{jt}(\vs_i)$, to the latent process $Z_t(i)$ with a normal cumulative density function (CDF) link, $\Phi$, and cutpoints $\lambda_j$; $\sum_{j=1}^{j} p_{jt}(\vs_i) = \Phi(\lambda_j - Z_t(i))$.
 For identifiability reasons, the first cutpoint, $\lambda_1$, is set equal to 0, and the last cutpoint, $\lambda_J$ set equal to infinity, with the remaining cutpoints learned.


In the original \cite{Albert1993BayesianData} framework, the latent process, $Z$, is assumed to vary only due to covariates. 
Here, we consider spatio-temporal covariates that can affect the probability of state transition, but also allow the state transition probabilities to depend on a latent low-rank spatio-temporal dynamic process to capture dynamics that are not well-specified by the neighbor-based covariates \citep[e.g., see][]{cressie2015statistics}.
In particular, we specify the latent process by 
\begin{equation} \label{eq:Z}
    \vZ_t = \vX_t \vbe + \vH \vY_t + \vep_t,
\end{equation}
where $\vZ_t \equiv [Z_t(1),\ldots,Z_t(n)]'$, $\vX_t$ is an $n \times p$ matrix of p covariates corresponding to each spatial location at time $t$ (see below), $\vbe$ are associated regression coefficients, $\vY_t$ is an $r$-dimensional dynamic process (where $r << n$), $\vH$ is an $n \times r$ spatial basis function matrix that maps this low-rank process to the $n$ spatial locations, and $\vep_t$ is a zero-mean error term constrained to be uncorrelated with variance 1, $\vep_t \sim N(\vzero,\vI)$, as in \cite{Albert1993BayesianData}.

Note that the data augmentation approach considered here is just one method to model ordinal data (e.g., see \cite{SCHLIEP20151} for an overview of additional methodologies for spatial data). For example, the P\'{o}lya Gamma approach of \cite{polson2013bayesian} has proven to be an efficient approach for Bayesian modeling of categorical data. In our case, 
the use of the data augmentation scheme from \cite{Albert1993BayesianData} was motivated by our consideration of a single unobserved latent process, $\vZ_t$. As shown in Appendix \ref{app:Polya}, an implementation of our model from the P\'{o}lya Gamma perspective provides essentially equivalent inference and prediction as the \cite{Albert1993BayesianData} approach presented here.

\subsubsection{Latent Spatio-Temporal Dynamics}
At the next level of the model hierarchy, we specify the evolution of the dynamic process $\vY_t$ in (\ref{eq:Z}) as a vector autoregression
\begin{equation}\label{eq:Yevolve}
\vY_t = \vM \vY_{t-1} + \veta_t,
\end{equation}
where $\veta_t \sim N(\vzero,\vQ)$ and $\vM$ is an $r \times r$ transition operator.
As discussed in \cite{wikle2019spatio}, it is important to allow the transition operator to be unstructured to capture realistic spatio-temporal dynamics (e.g., transient growth).
This is facilitated computationally by the low-rank basis representation, which is often considered in spatio-temporal dynamic models given that the underlying dynamics typically exist on a lower-dimensional manifold than that on which it is modeled \citep[e.g.,][]{wikle2019spatio}.

We discuss implementation issues associated with the choice of covariates and basis functions in \ref{subsec:ModImp}. In addition, the next level of the model hierarchy requires specification of prior distributions for \{$\gamma_j: j = 2, \dots J-1$\} parameters, $\vbe$, $\vM$, $\vQ$, and $\vY_0$.
These are presented in Section \ref{subsec:ModImp} as well.

\subsection{Model Implementation}\label{subsec:ModImp}

This section discusses implementation choices associated with the local covariates in (\ref{eq:Z}), the choice of spatial basis functions in (\ref{eq:Z}), the prior distributions for model parameters, the MCMC algorithm, and our approach to model evaluation.

\subsubsection{Specification of Local Covariates}

\begin{figure}[H]
    \centering
    \includegraphics[width=0.75\textwidth]{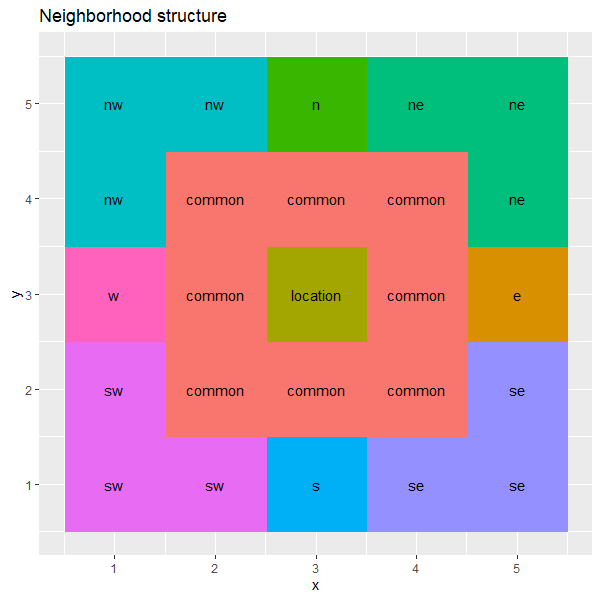}
    \caption{Neighborhood structure used for the S5 fire considered in Section 4; if the wind is from the southwest (sw) direction with x and y wind vectors $>2$m/s then the neighborhood includes the three cells from the sw. Likewise, if the wind is from the south (s) with a wind speed in the y direction $>2$ m/s and the x wind vector $\leq2$m/s, then only an additional cell from the south is included in the structure.}
    \label{fig:windNeigh}
\end{figure}

The latent $\vZ_t$ process in (\ref{eq:Z}) is a function of local time-varying covariates, $\vX_t$, whose specification is motivated by empirical thermodynamic relationships that are sometimes used to characterize fire spread.
Specifically, \cite{RothermelRAFuels.} proposed perhaps the most-used thermodynamic relationship for the spread of fires \citep[see also][for further discussion]{AndrewsTheExplanation}: 
\begin{equation}\label{eq:Rothermel}
    R = \frac{I_R \xi (1 + \theta_w + \theta_s)}{\rho_b \zeta Q_{ig}},
\end{equation}
where the rate of fire spread, $R$ (ft/min), is defined as a function of reaction intensity ($I_R$), larger-scale propagating flux ratio (i.e., a long-range wind factor $\theta_w$), a slope factor ($\theta_s$), and scaled by properties of the fuel ($\rho_b, \zeta, Q_{ig}$).
Wind speed influences the size and shape of the neighborhood to match the physical properties of the Rothermel equation. Rather than use this equation directly, as it is usually used in fire modeling,  we utilize this relationship to modify the neighborhood structure of our CA model dynamically in time as a function of wind speed.
That is, when the wind speed in either the east-west or north-south direction is greater than $2 m/s$, then the wind expands the neighborhood in the corresponding direction, as shown in Figure  \ref{fig:windNeigh}.
This novel dynamic neighborhood structure, where the neighborhood can change at each time point, allows for the model to capture larger-scale dynamics, as in \cite{Zhang2021AModels}, but is motivated by well-known physical and thermodynamic relationships.
Thus, the matrix $\vX_t$, an $n$ by $p = 3$ matrix, consists of the state of the cells in their dynamic neighborhood.
The use of three states is again motivated by the Rothermel equation because the heat flux into a cell is a function of the fuel property and an unburnt cell is fundamentally different from a burnt cell due to the change in fuel properties of the cell.
Denoting $\mathcal{N}_{k,t}$ to be the neighborhood structure of cell k at time $t$, we can then write $X_{k,t,1} = \sum_i  I(S_{t-1}(s_i) = 1), i \in \mathcal{N}_{k,t}$, $X_{k,t,2} = \sum_i  I(S_{t-1}(s_i) = 2), i \in \mathcal{N}_{k,t}$, and $X_{k,t,3} = \sum_i  I(S_{t-1}(s_i) = 3), i \in \mathcal{N}_{k,t}$,
where $S_{t-1}(s_i)$ is the state at location i (which is a neighbor of cell k) at the previous time.

\subsubsection{Specification of Spatial Basis Functions} \label{subsec:BasisFn}
The reduced-rank dynamic process $\vY_t$ in (\ref{eq:Z}) relies on $r$ spatial basis functions in the $n \times r$ matrix $\vH$. There are many choices for such basis functions \citep[e.g.,][]{wikle2019spatio}.
Here, we consider a constructed empirical orthogonal function (EOF) spatial basis.
As discussed in \cite{cressie2015statistics}, EOFs can be good spatial basis functions for dynamical processes because they provide an optimal (from the Karhuen-Lo\`eve decomposition) low-rank spatial dimension reduction with realistic covariance structure (note, sensitivity to the choice of basis functions is briefly discussed in Section \ref{sec:S5}). 
The calculation of EOFs requires time replicates. 
Although we have time replicates for fire data, if one is predicting the spread outside the range of the fire front, the basis may not provide adequate coverage of unburnt areas.
We address this with a novel constructed EOF algorithm.

The constructed EOF approach requires that we initially simulate a plausible realization of fire spread. 
Specifically, the final state of the fire at the last observation time $T$ is used as an input to a simple model.
The simple model, $f_{sim}$, can then be run for the desired number of time steps, $\tau$, and the state of the fire can be predicted using the simulation. 
The simulation model predicts the probability of each state at time $T + 1$, $T + 2$, ..., $T + \tau$ and the most probable state used to evolve the fire.
Once a collection of predicted states is computed, each cell can randomly be assigned a temperature value, $\vZ_t : t = T+1, \ldots$, from that same state using $g_{sim}$.  
This approach for constructing EOFs is given in Algorithm \ref{alg:eof} and the details of the specification of $f_{sim}$ and $g_{sim}$ is further discussed in Appendix \ref{Ap:EOFConstruct}. 

Importantly, the simulation model, $f_{sim}$, is not meant to be an ideal simulator for a particular fire. 
The use of the simplified simulator is simply to evolve the fire front sufficiently to allow the EOF basis to cover unburnt regions in the actual CA model as presented in Section \ref{sec:propMethod}.

The choice of EOFs as a basis set is subjective, but it does have some advantages compared to other bases, such as spatial bisquare bases \citep{wikle2019spatio}.
The primary advantage of using the EOF basis functions is that they are guaranteed to capture the spatial dependence most efficiently for a given number of basis functions.
Having fewer spatial basis function coefficients is essential in our application due to the estimation of the latent dynamical process parameters (i.e., $\vM, \vQ$) given a relatively few number of time points. See Appendix \ref{Ap:BasisSensitivity} for a demonstration of the sensitivity of our model to the specification of a low-rank bisquare basis set. 
\begin{algorithm}
\caption{Construction of the EOFs}\label{alg:eof}
\begin{algorithmic}
    \Require $\vS: \text{n spatial locations at T time points}$
    \Require $\vZ_t(s = 1), \vZ_t(s = 2), \vZ_t(s = 3)$
    \While{$t \leq \tau$}
        \State $\vS_{T + t} \gets f_{sim}(\vS_{T + t - 1})$
        \State $\vZ_{T + t} \gets g_{sim}(\vS_{T + t})$
    \EndWhile  \\
    $\vH \gets EOF(\vZ_{1, 2, ..., T + \tau})$
\end{algorithmic}
\end{algorithm}

\subsubsection{Prior Specification and MCMC Algorithm}

To complete the Bayesian hierarchical model, non-informative priors are assigned to the model parameters.
The local covariate effects, $\vbe$, are assigned independent $N(0,5)$ priors.
Elements of the evolution matrix, $\vM$, are assigned independent   $N(0, 2)$ priors, and the inverse innovation covariance matrix is assigned an inverse Wishart prior, $\vQ^{-1} \sim W((\vI_r r)^{-1}, \vI_r)$.
The cutoffs, $\lambda_j$, are given improper flat priors, as in \cite{Albert1993BayesianData}.

Once the local covariates, spatial basis functions, and prior distributions are chosen, inference and prediction is accomplished through Markov chain Monte Carlo (MCMC) methods as outlined in Algorithm \ref{alg:samp} and presented in detail in Appendix \ref{ApSamp}.

\begin{algorithm}
\caption{General Sampling algorithm, see Appendix \ref{ApSamp} for full details.}\label{alg:samp}
\begin{algorithmic}
    \Require $\vH$, $\vX_t$, $\vZ_t$
    \Procedure{Sampling}{}
    \While{sampling == TRUE}
        \State $\vY_t| \cdot  \sim N(\cdot)$
        \State $\vQ^{-1}| \cdot \sim Wis(\cdot)$
        \State $\vbe| \cdot \sim N(\cdot)$
        \State $\vZ_t| \cdot \sim N(\vX_t \vbe + \vH \vY_t, \vI)$
        \State $\lambda| \cdot \sim Unif(\max [\max(Z_i: S_i = j, \lambda_{j-1}], \min [\min (Z_i:S_i = j +1, \lambda_{j+1}])$
    \EndWhile
    \EndProcedure
\end{algorithmic}
\end{algorithm}

\subsubsection{Model Evaluation}
The model is evaluated using two different forecast verification metrics that are specifically designed for categorical data -- the ranked probability score (RPS) and Gilbert Skill Score (GSS), as discussed in \cite{jolliffe2012forecast}.

GSS is a measure based on the $2 \times 2$ contingency table of the forecasted states.
Our model has 3 predicted states, but we can compress these into two classes (``burning'' and ``not burning'') so that the contingency table can be rewritten as:
\begin{table}[H]
    \centering
    \begin{tabular}{l | c c}
         & Reference & \\
         \hline
         Prediction & Burning & Not Burning \\
         \hline
         Burning & a & b \\
         Not Burning & c & d\\
    \end{tabular}
    \caption{$2 \times 2$ Contingency table for computing GSS}
    \label{tab:GSSCont}
\end{table}
\noindent GSS is then calculated by:
\begin{align*}
    \text{GSS} & = \frac{a - \frac{(a+c)(a+b)}{n}}{a + b + c - \frac{(a+c)(a+b)}{n}},
\end{align*}
where the term $(a+c)(a+b)/n$ can be considered to be the expected number of true predictions due to chance.
A larger GSS indicates a better model, with 1 being a perfect prediction. GSS has been used for forecast evaluation in meteorology, climatology, and related fields  \citep[e.g.,][]{Schaefer1990,Stephenson2000}.

In addition to being able to predict a future state, our model is able to forecast the probability of each state.
Thus, it is appropriate to consider an evaluation approach that considers the probability.
The RPS \citep[e.g.,][]{EpsteinRPS, MasonRPSS,Ferro2008} is a classic scoring rule that considers the probability of each cell being state $j$.
The RPS for a single prediction location is calculated by:
\begin{align*}
    \text{RPS} & = \frac{1}{J - 1} \sum_{j = 1}^{J}[(\sum_{k = 1}^{j}p_k - \sum_{k = 1}^{j} I_k)]^2,
\end{align*}
where J is the number of categories, in this case 3.
The RPS for each forecast location is calculated and the average value across all locations is reported.
A lower RPS indicates a better model with 0 indicating a perfect prediction.


\section{Simulation Experiment} \label{sec:Sim}

The purpose of this simulation is to ensure that the ordinal categorical CA model can retrieve the true probabilities of transitioning.
Thus, we consider a simulated fire evolving in time that is categorized into three states; unburnt, burning, and burnt.
The simulated fire considers a spatial domain of $20 \times 30$ cells with 45-time points - a similar size to the real-world application presented below.
The rule for the fire propagation (i.e., from state unburnt $\to$ burning) was a function of the number of burning Moore neighbors, see Table \ref{tab:simResult} for the transition probabilities based on the number of burning neighbors.
This rule roughly approximates the physical propagation of a fire front as given by the Rothermel equation (Eq. \ref{eq:Rothermel}) as discussed in Section \ref{sec::Method-model}.
The Bayesian hierarchical model described in Section \ref{sec:propMethod} was fitted to this dataset.
There was an intermittent wind term in the positive y direction, so the neighborhood structure was a function of this wind term, using the neighbors as shown in Figure \ref{fig:windNeigh}.
All state transitions were a function of this dynamic neighborhood structure, so the latent $\vY_t$ process was omitted.
The MCMC was run for 10,000 iterations, with the first half discarded as burn-in.

The first 40 observations from the simulated dataset were used to train the model, and the in-sample prediction, as seen in Table \ref{tab:simResult}, shows that the method can retrieve the true probabilities of state transitions.

\begin{table}[H]
    \centering
    \begin{tabular}{l|c|c|c}
         Metric & Unburnt & Burning & Burnt  \\
         \hline
         Prediction Correct & 15193 (81) & 4035 (226) &  3796 (69) \\
         GSS & 0.975 & 0.897 & 0.923
    \end{tabular}
    \caption{Simulation Results - Prediction Correct is the number of correct (incorrect) by count (e.g., the model correctly predicted state ``unburnt'' 15,193 times and predicted state ``unburn'' 81 times, but the truth was otherwise). The simulation had an RPS of 0.022 (versus the na\"{i}ve RPS of 0.747 from equal probabilities for each state)}
    \label{tab:simResult}
\end{table}

The overarching goal of this CA model is to learn the probabilities of the three-state fire spread model and use these learned rules to forecast the fire spread.
Table \ref{tab:simBurning} shows that the CA approach presented here can correctly recover the probability of state transition based on the local neighborhood.
The model can also be used for forecasting. 
The next five time points beyond the training period were used to test forecast performance, and the RPS was 0.100 versus the na\"{i}ve RPS of 0.579 based on equal probabilities.
Incorporating the CA modeling framework within a Bayesian inference paradigm allows us to quantify the uncertainty of the transitions between states.
The HPDs in Table\ref{tab:simBurning} demonstrate the uncertainty associated with the transition probabilities; thus, forecasted and predicted states from this model include uncertainty measures that are unavailable in traditional deterministic CA models. 

\begin{table}[H]
    \centering
    \begin{tabular}{|| c | c | c | c ||}
    \hline
         \# unburnt & \# burning &  Prob Burning & HPD \\
         \hline 
         0 & 8 & 1.0000 & \textbf{(1,1)} \\
         1 & 7 & 1.0000 & \textbf{(1, 1)} \\
         2 & 6 & 0.9999 & \textbf{(0.9999, 1)} \\
         3 & 5 & 0.9994 & \textbf{(0.9989, 0.9999)} \\
         4 & 4 & 0.9641 & \textbf{0.9533, 0.98163)} \\
         5 & 3 & 0.6368 & \textbf{(0.6127, 0.7188)} \\
         6 & 2 & 0.1357 & \textbf{(0.1295, 0.1889)} \\
         7 & 1 & 0.0054 & \textbf{(0.0047, 0.0109)} \\
         8 & 0 & 3.167e-05 & \textbf{(1.974e-05, 1.165e-04)} \\
         \hline
    \end{tabular}
    \caption{The true probability of advancing from state unburnt to burning based solely on the number of neighbors burning and unburnt. The truth from the simulation and the recovered learned probabilities from the model are presented as 95\% HPD intervals, with bold intervals indicating those which contain the true parameter.}
    \label{tab:simBurning}
\end{table}

\section{S5 Fire Example} \label{sec:S5}

The real-world data for this application comes from the RXCadre series of experimental burns from Florida, USA.
An infrared camera recorded the area's temperature, and local weather conditions were recorded \citep{RXcadreExperiment}.
The data were originally $240 \times 320$ resolution with 105-time points. For the analysis presented here, the data were  averaged onto a 24x32 grid  with the mean temperature of the pixels within each larger cell used as data. 
These temperatures were then categorized into three states: burnt, burning, and unburnt.
The criteria for the classification were based on the temperature of the cell, with progression from unburnt to burning after the temperature crossed a threshold above the background average of 300K.
After a burning cell returned to the average background temperature of 300K, the cell was considered to have transitioned to a burnt state.
Using discrete states instead of modeling the temperature allows the model considered here to be applied to a broader range of data sources (e.g., those that just present the fire front boundary at a given time).

\begin{figure}[H]
    \centering
    \includegraphics[width=\textwidth]{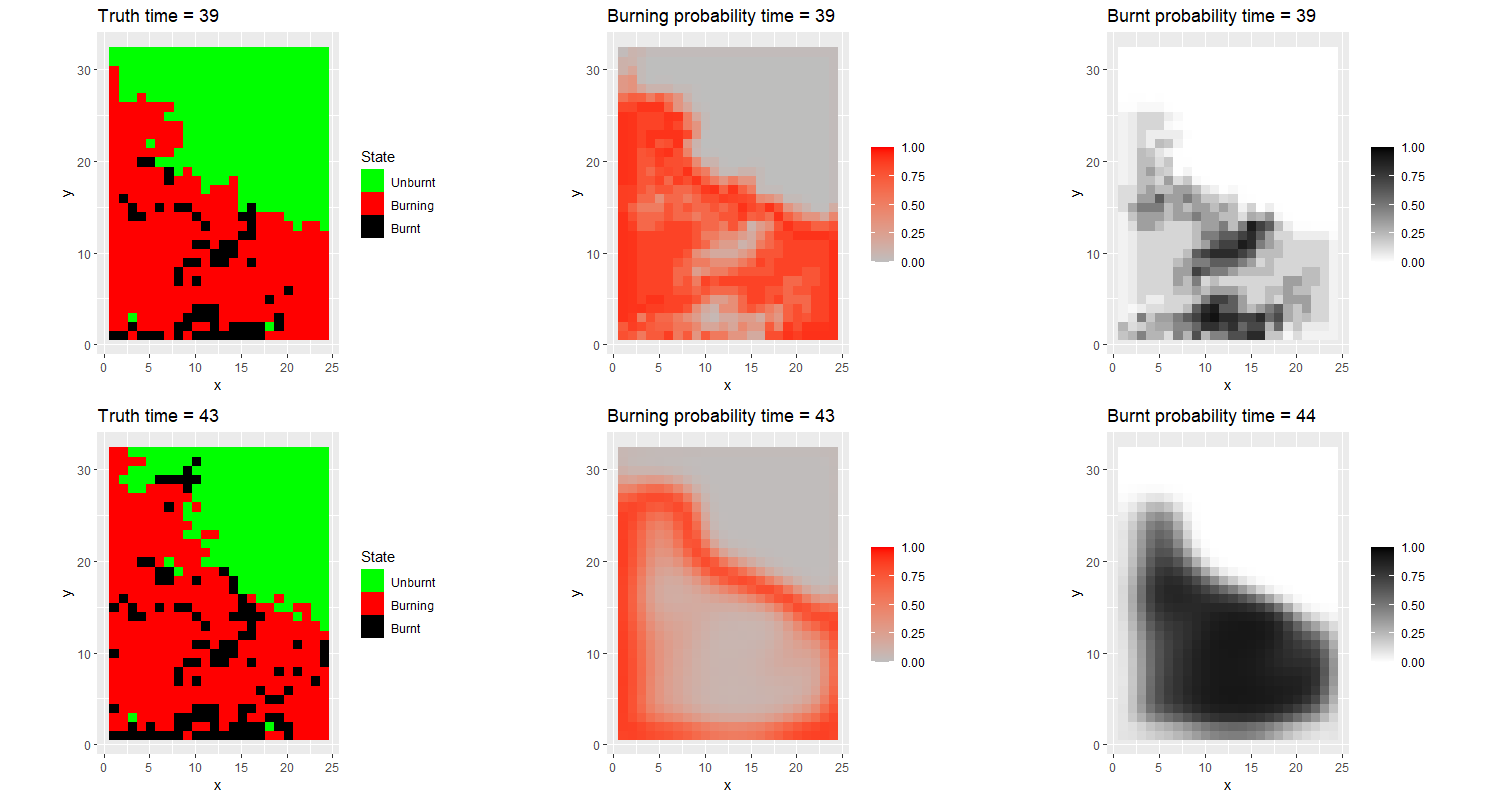}
    \caption{S5 fire forecasting using covariates only with no latent temporal process ($\vY_t$). The use of only local covariates is insufficient to explain the spread of the fire. The left side is the truth at the beginning and end of the forecast period. The middle panel shows the mean predicted probability of the cell being in the burning state. The right side of the figure shows the mean predicted probability for the burnt state. The top of the panel shows the first forecast step at timepoint 39, and the bottom of the panel is the five timepoint forecast at timepoint 43.}
    \label{fig:S5foreCov}
\end{figure}

The first 39 observations were used to train the model, and then the trained model was used to forecast 5 time points forward.
The MCMC sampler was run for 10,000 iterations, with the first 5,000 discarded as burn-in. 
First, a model using only the covariates $\vX_t$, with no latent temporal process was fitted, and the results are shown in Figure \ref{fig:S5foreCov}.
The model can capture the fire's growth in the positive x and y direction, but the local covariates are insufficient to model realistic spread.
For example, the model predicts the transition from the burning state to the burnt state to be more rapid than the observations, as seen in the right side of Figure \ref{fig:S5foreCov}.

\begin{figure}[H]
    \centering
    \includegraphics[width=\textwidth]{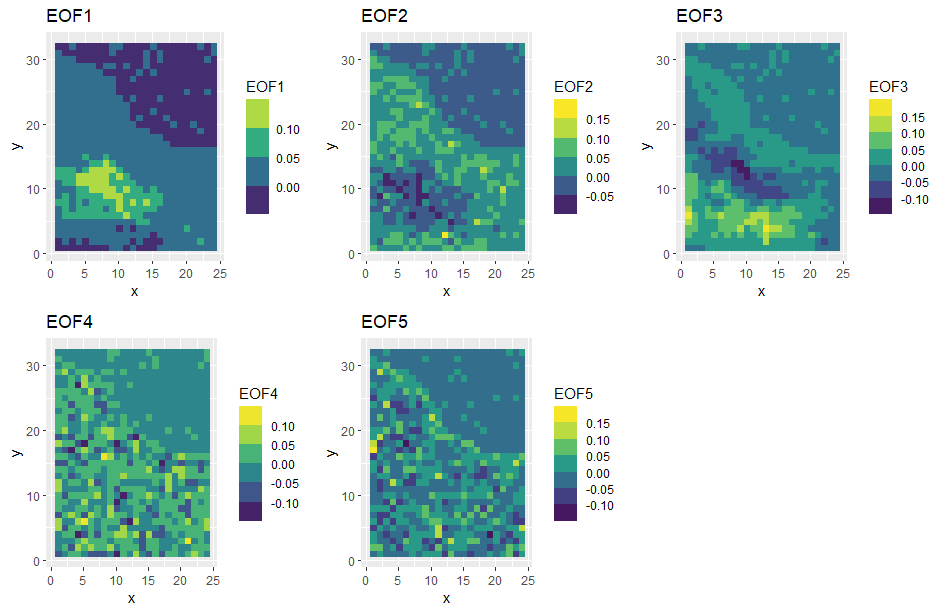}
    \caption{The first 5 EOFs after forecasting the fire using the simulation model. These are the basis functions used to link the latent temporary evolving $Y$ process back to the original spatial domain.}
    \label{fig:EOFval}
\end{figure}

The limitation illustrated by only considering the covariates to model fire spread can be addressed by the addition of the latent dynamical process $\vY_t$ linked by the spatial basis functions.
The first 5 EOFs were constructed following the procedure in Section \ref{subsec:BasisFn}; Figure \ref{fig:EOFval} shows the general structure.

\begin{figure}[H]
    \centering
    \includegraphics[width=\textwidth]{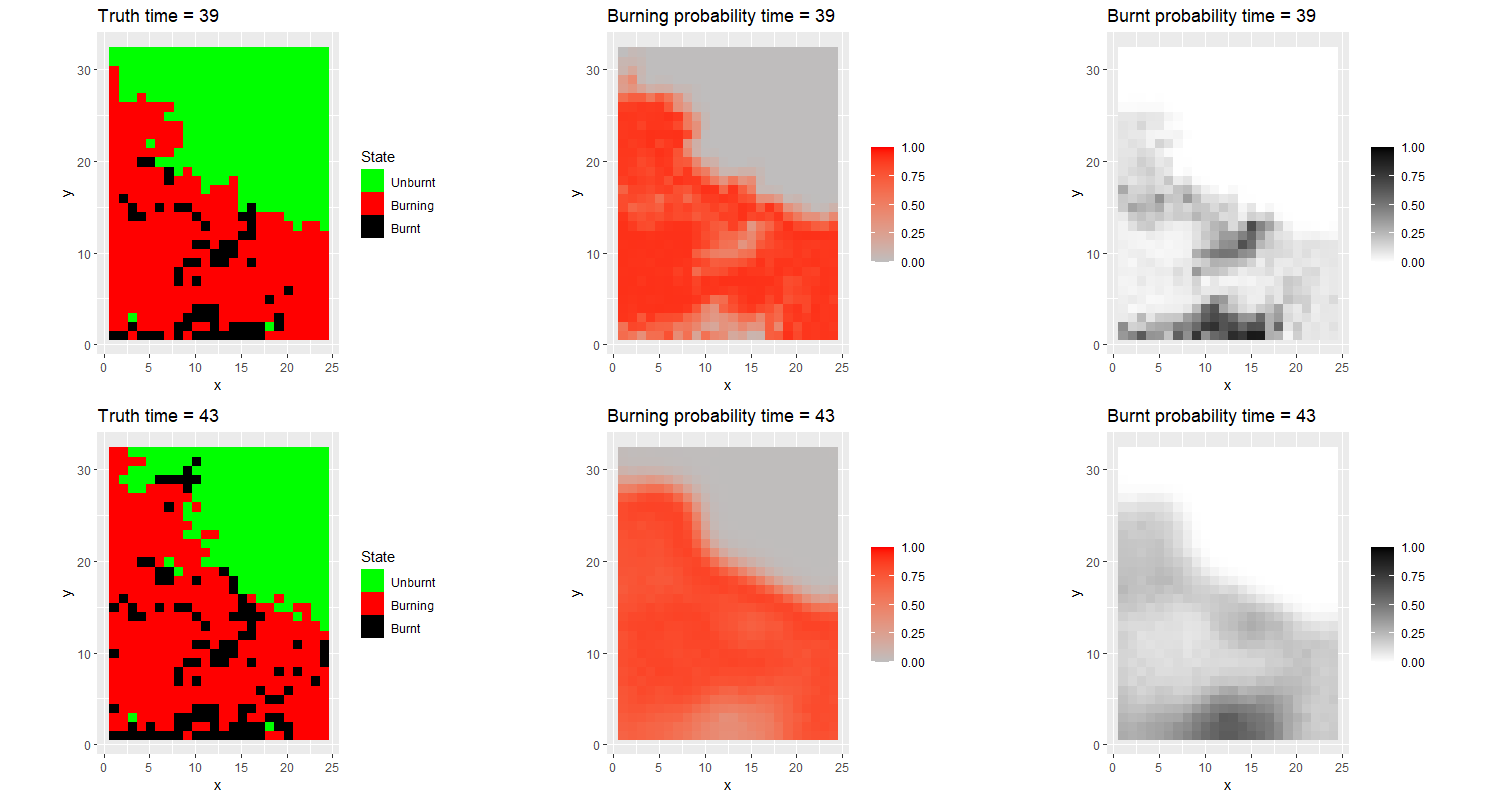}
    \caption{S5 fire forecasting using the proposed method with the addition of the temporally evolving latent $\vY_t$ dynamic process. The left panel shows the true states, and the middle panel shows the mean prediction probability of the burning states. The right panel is the mean probability of burnt state, which is significantly reduced when compared to Figure \ref{fig:S5foreCov}.}
    \label{fig:S5foreEOF}
\end{figure}

\begin{figure}[H]
    \centering
    \includegraphics[width=\textwidth]{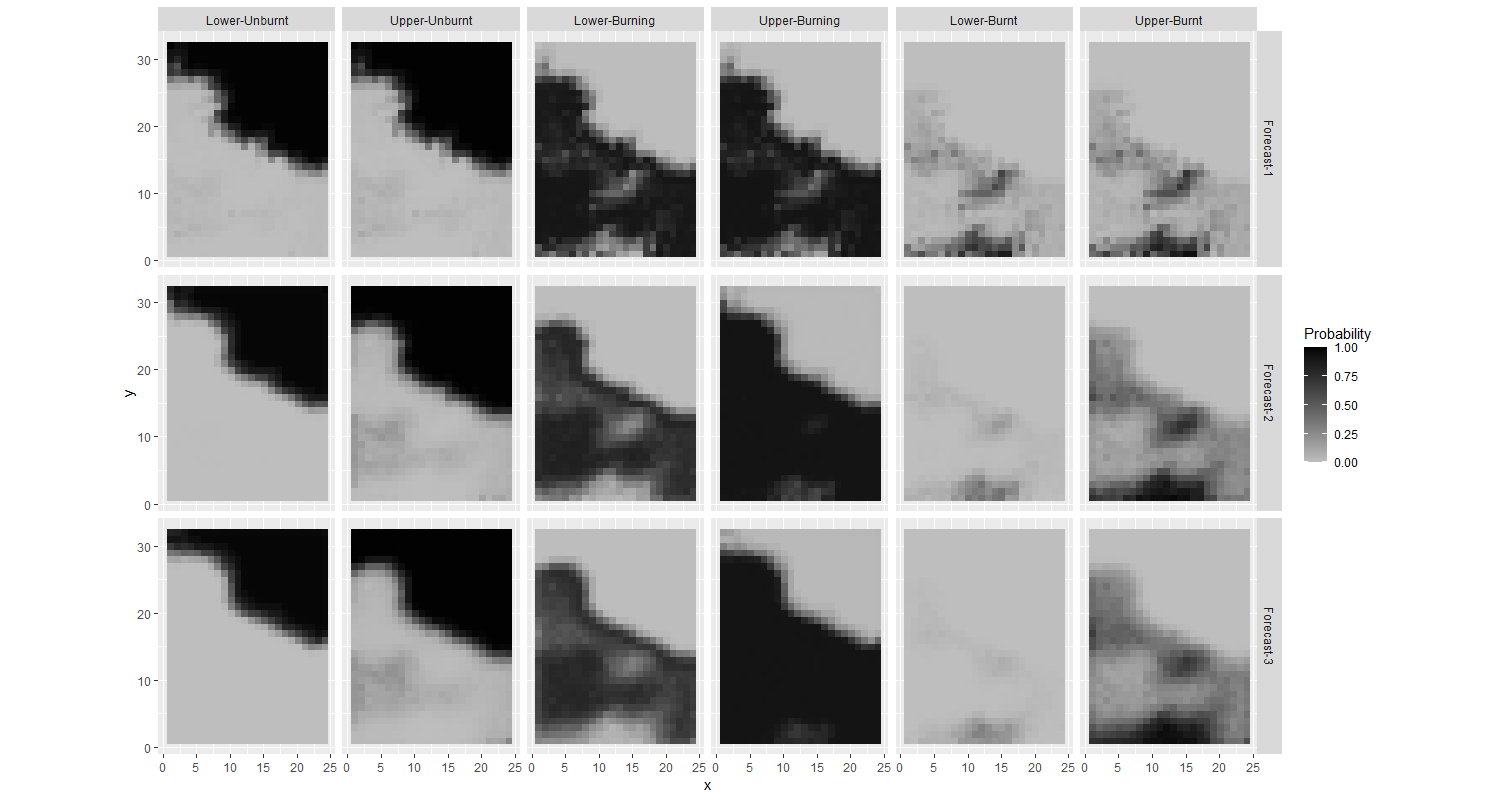}
    \caption{HPD intervals for the first 3 time forecasts for the three possible states. The top row is for the first forecast time period. The first two panels in each row show the lower and upper values of the HPD for the unburnt state, the middle two panels are the bounds for the burning state, and the right two panels are the bounds for the burnt state. The subsequent rows are for the next two forcast time periods. As the length of the forecast increases, all of the HPDs become wider, as expected.}
    \label{fig:facetForecast}
\end{figure}

As shown in Table \ref{tab:S5result} and visually in Figure \ref{fig:S5foreEOF}, adding the latent dynamical process $\vY_t$ on the expansion coefficients from  the first 5 EOFs leads to an improvement in prediction accuracy.
The fire in Figure \ref{fig:S5foreEOF} is forecast to progress more rapidly in the positive x and y directions, providing a better match to the truth than the forecast with only local covariate effects - this is further supported by the model scores in Table \ref{tab:S5result}.
The GSS increases for each category with the largest improvement in capturing the growth of the fire. 
The RPS decreased, showing an improved model fit with the addition of the latent dynamical process.
Figure \ref{fig:facetForecast} demonstrates the strength of using a Bayesian hierarchical model.
The model is able to provide bounds on the state estimation for each forecast time.
The results in Appendix \ref{ApS7} investigate an additional fire with different fire spread characteristics, but still from the RXCadre experiment. This example also demonstrates that the model with the latent spatio-temporal dynamics performs better than the covariate-only model.

\begin{table}[H]
    \centering
    \begin{tabular}{l|c|c|c}
         Metric & Unburnt & Burning & Burnt  \\
         \hline
         \textbf{Covariates Only} & & & \\
         Prediction Correct & 1332 (76) & 1074 (309) &  260 (789) \\
         GSS & 0.778 & 0.248 & 0.127 \\
         \hline
         \textbf{Simulated EOFs} & & & \\
         Prediction Correct & 1341 (77) & 1821 (480) &  84 (37) \\
         GSS & 0.8111 & 0.70 & 0.185 
    \end{tabular}
    \caption{S5 fire results - Prediction Correct is the number of correct (incorrect) forecast states. The covariate-only model had an RPS of 0.354 (versus the na\"{i}ve RPS of 0.712), and the latent dynamic model had an RPS of 0.212}
    \label{tab:S5result}
\end{table}

\section{Conclusion} \label{sec:Conc}

The Bayesian hierarchical cellular automata model presented here offers a few key benefits.
A three-state model is more justifiable from a physical perspective than a two-state model with only burning and unburned states. 
This is because burnt cells would have smaller addition to the heat flux in an adjacent cell than a model that only allows for the transition from unburnt to burning.
The CA model prsented here could incorporate local covariates such as (but not limited to) wind speed, by dynamically varying the definition of neighbors.
In addition, the model matrix, $\vX_t$, can easily be expanded to include terms measuring the cell's fuel if the fire spreads over a non-homogeneous region or include slope properties - as both are terms in the Rothermel equation.
Importantly, adding the unobserved latent dynamic process $\vY_t$ explains some of the dynamics not captured by the local neighborhood structure, and this latent process is linked using a novel construction of low-rank EOF basis functions.
Finally, the model can easily be used to forecast a fire's spread and quantify prediction uncertainty, and this uncertainty of the prediction can be allowed to propagate throughout the length of the forecast.

Future extensions of the method could explore different definitions of neighbors.
In the model presented here, the neighborhood structure was novel in that it was defined based on the Rothermel equation, but data-driven methods could be used to learn this structure.
 In addition, adding a fire spotting process to the model could aid in capturing some of the dynamic fire front shifts that happen in large-scale real-world fires.
Mechanisms to allow for missing data need to be considered in cases where the data is not reliably recorded on a regular spatial and temporal scale.

\newpage
\appendix
\section{MCMC Sampler} \label{ApSamp}

The full-conditional distributions for our MCMC algorithm are presented below.
\subsection{Priors}
\begin{align*}
    [\vbe] & \sim N(\vmu_b, \vSig_b) \\
    [\vY_0] & \sim N(\vmu_0, \vSig_0) \\
    [\vQ^{-1}] & \sim W( (\vnu_Q C_Q)^{-1}, \vnu_Q) \\
    [\vm] & \sim N(\vmu_m, \vSig_m) \\
\end{align*}    
Hyperparameters were fixed with $\vmu_b = \vmu_m = \vmu_0 = 0$, $\vSig_b = \vSig_m = diag(2)$, $\nu_Q = 1$, and $\vSig_0 = diag(5)$
\subsection{Sampler}

\begin{equation*}
    \vZ_t = \vX_t \vbe + \vH \vY_t + \vep
\end{equation*}

\begin{enumerate}
    \item $\vY_0 \sim N(V_0 a_0, V_0)$
    \begin{itemize}
        \item $V_0 = (\vM' \vQ^{-1} \vM + \Sigma_0 ^{-1})^{-1}$
        \item $a_0 = \vM' \vQ^{-1} \vY_1 + \Sigma_0^{-1} \mu_0$
    \end{itemize}
    \item $\vY_t \sim N(V_t a_t, V_t)$
    \begin{itemize}
        \item $V_t = (\vH' \vR^{-1} \vH + \vQ^{-1} + \vM' \vQ^{-1} \vM)^{-1}$
        \begin{itemize}
            \item Note this is where the computation savings come in. We can invert an r x r matrix faster than an n x n.
        \end{itemize}
        \item $a_t = \vH' \vR^{-1} (\vZ_t - \vX_t \vbe) + \vQ^{-1} \vM \vY_{t-1} + \vM' \vQ^{-1} \vY_{t+1}$
    \end{itemize}
    \item $Y_{t = T} \sim N(V_T a_T, V_T)$
    \begin{itemize}
        \item $V_T = (\vH' \vR^{-1} \vH + \vQ^{-1})^{-1}$
        \item $a_T = \vH' \vR^{-1} (\vZ_t - \vX_t \vbe) + \vQ^{-1} \vM \vY_{T-1}$
    \end{itemize}
    \item $\vQ \sim W(Q_1, Q_2)$
    \begin{itemize}
        \item $Q_1 = (\sum (\vY_t - \vM \vY_{t-1})(\vY_t - \vM \vY_{t-1})' + \nu_Q  C_Q)^{-1} $
        \item $Q_2 = \nu_Q + T$
    \end{itemize}
    \item $vec(\vM) \sim N(V_m a_m , V_m)$
    \begin{itemize}
        \item $V_m = ((\vY'_{0:T-1} \otimes \vI_n)'(\vI_T \otimes \vQ)^{-1}(\vY'_{0:T-1} \otimes \vI_n) + \Sigma_m^{-1})^{-1}$
        \item $a_m = (\vY'_{0:T-1} \otimes \vI_n)'(\vI_T \otimes \vQ)^{-1} vec(\vY_{1:T}) + \Sigma_m^{-1} \mu_m$
    \end{itemize}
    \item $\beta^i \sim N(V_b a_b, V_b)$
    \begin{itemize}
        \item $V_b = (\vX' \vI \vX +\Sigma_b^{-1})^{-1}$
        \item $a_b = (vec(\vZ - \vH \vY)' \vI_{nT,nT} \vX + \mu_b \Sigma_b^{-1})'$
    \end{itemize}    
    \item $\vZ | S \sim N(\vX \vbe + \vH \vY, \vI)$
        \begin{itemize}
            \item Truncated on the left and right by $\lambda_{j-1}, \lambda_{j}$ for $S_i = j$
            \item Note for identifiability reasons $\lambda_1 = 0$. $\lambda_0 = -\infty$ and $\lambda_{J+1} = \infty$
        \end{itemize}
    \item $\lambda_i \sim \text{Unif}(\max [\max(Z_i: S_i = j, \lambda_{j-1}], \min [\min (Z_i:S_i = j +1, \lambda_{j+1}])$
    
\end{enumerate}

\section{EOF Construction}\label{Ap:EOFConstruct}

The method to construct the EOF is as follows.
From time $t=1$ to the end of the training period, time $t = T$, a multinomial model, $f_{sim}$, was fit with only the cell's neighbors states as covariates.
This model was then used to forecast the most probable state of the fire up to time $T + \tau$, where $\tau$ is the forecast length.
To finalize the construction of the EOF, the temperature of each cell needs to be known.
From time $t = 1$ to $t = T$ this is given from the data, but for time $T + 1$ to $T + \tau$ these need to be simulated.
Using the temperature at the training time points, the simulated temperature was randomly sampled from this distribution, $g_{sim}$.
For example, if a cell was forecasted to be state ``burning'', the temperature was randomly sampled (imputed) from the distribution of all cells that were state ``burning'' from time $t = 1$ to $t = T$.

\section{S7 fire} \label{ApS7}

\begin{figure}[H]
    \centering
    \includegraphics[width=\textwidth]{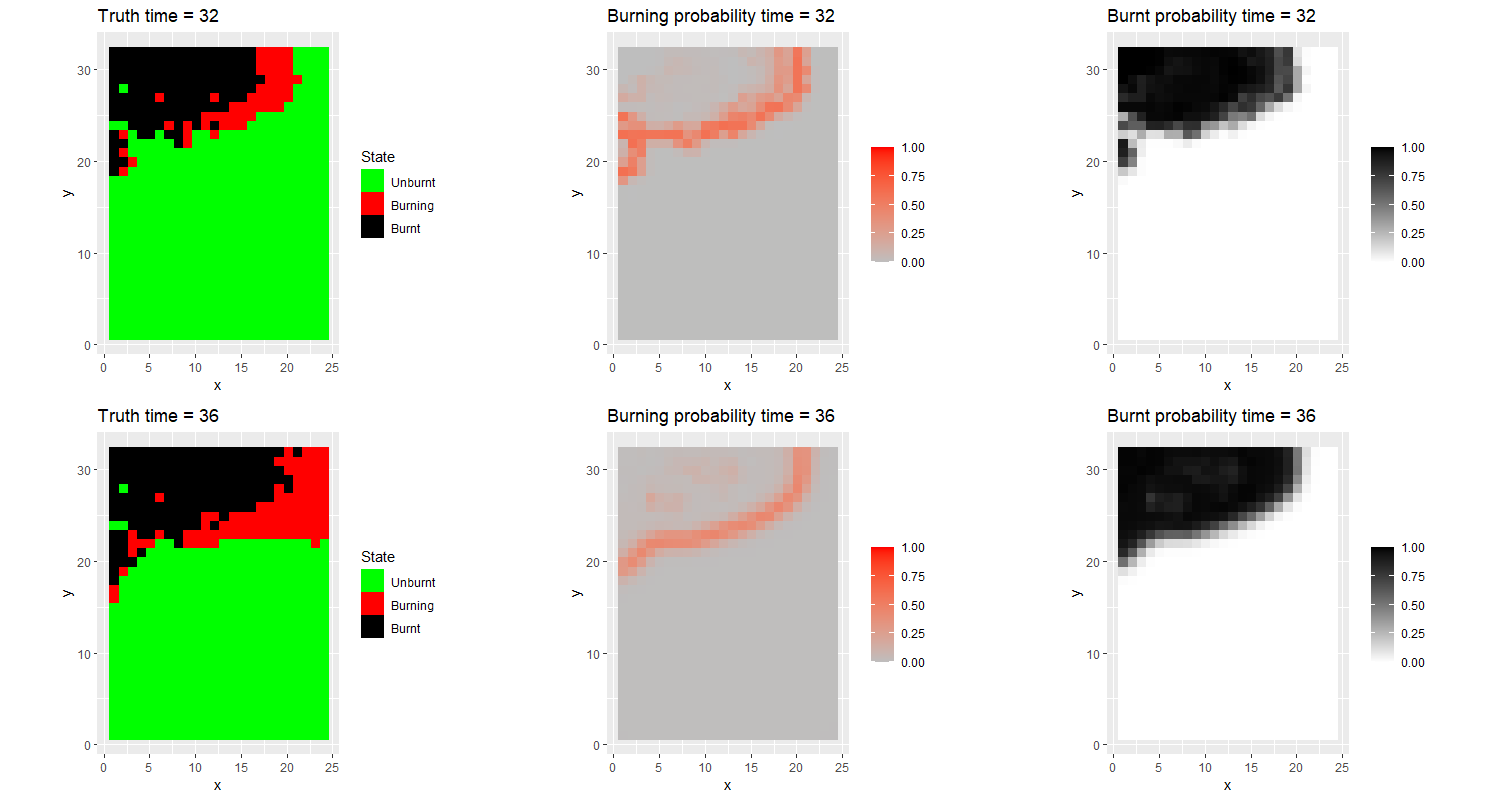}
    \caption{S7 simulated EOF forecasts}
    \label{fig:S7fore}
\end{figure}

In addition to the S5 fire discussed in Section \ref{sec:S5}, we consider another fire from the RXCadre experiment that demonstrated different characteristics than the S5 fire
This fire, S7, had a fire front that was more steadily burning with the fire lacking some of the large jumps in fire boundary that were experienced in the S5 fire.
In addition, the cells transitioned to the burnt state more rapidly than in the S5 fire.
As can be seen in Figure \ref{fig:S7fore}, there was some exceptionally rapid growth that occurred that the model was unable to capture.
However, as demonstrated in Table \ref{tab:S7result} the addition of the simulated EOFs coupled with the dynamic $\vY_t$ process was able to outperform the model with only local covariates.
The GSS was higher in all categories which indicates a better fit and the RPS was lower which supports the same conclusion.

\begin{table}[H]
    \centering
    \begin{tabular}{l|c|c|c}
         Metric & Unburnt & Burning & Burnt  \\
         \hline
         \textbf{Covariates Only} & & & \\
         Prediction Correct & 2459 (16) & 0 (3) &  751 (311) \\
         GSS & 0.709 & 0.000 & 0.442 \\
         \textbf{Simulated EOF} & & & \\
         Prediction Correct & 2698 (205) & 60 (34) &  710 (133) \\
         GSS & 0.729 & 0.133 & 0.758
    \end{tabular}
    \caption{S7 fire results - Prediction Correct is the number of correct (incorrect) by count. The covariate model had an RPS of 0.282 and the simulated EOF model an RPS of 0.151 (versus the naive RPS of 0.733).}
    \label{tab:S7result}
\end{table}

\section{P\'{o}lya Gamma Augmentation} \label{app:Polya}

The use of the data augmentation scheme from \cite{Albert1993BayesianData} was motivated by the use of a single unobserved latent process, $\vZ_t$.
Other data augmentation strategies for categorical data have been proposed, such as \cite{polson2013bayesian}, which utilize a latent P\'{o}lya Gamma process for each category.
Both augmentation schemes produce similar results as demonstrated by considering the simulation from Section \ref{sec:Sim}.
The HPD intervals for the P\'{o}lya Gamma augmentation scheme are shown in Table \ref{tab:simBurningPolya} and are similar to the results in Table \ref{tab:simBurning}.

\begin{table}[H]
    \centering
    \begin{tabular}{|| c | c | c | c ||}
    \hline
         \# unburnt & \# burning &  Prob Burning & HPD \\
         \hline 
         0 & 8 & 1.0000 & (0.958,0.998) \\
         1 & 7 & 1.0000 & (0.975, 0.998) \\
         2 & 6 & 0.9999 & (0.985, 0.998) \\
         3 & 5 & 0.9994 & \textbf{(0.989, 0.997)} \\
         4 & 4 & 0.9641 & \textbf{0.957, 0.980)} \\
         5 & 3 & 0.6368 & \textbf{(0.621, 0.753)} \\
         6 & 2 & 0.1357 & \textbf{(0.093, 0.152)} \\
         7 & 1 & 0.0054 & (0.006, 0.012) \\
         8 & 0 & 3.167e-05 & \textbf{(0.000, 0.001)} \\
         \hline
    \end{tabular}
    \caption{The true probability of advancing from state unburnt to burning based solely on the number of neighbors burning and unburnt. The truth from the simulation and the recovered learned probabilities for the Polya Gamma data augmentation approach presented as 95\% HPD intervals, with bold intervals indicating those which contain the true parameter.}
    \label{tab:simBurningPolya}
\end{table}

The P\'{o}lya Gamma data augmentation scheme produced similar results to the Albert and Chib method while forecasting for a the same length.
The P\'{o}lya Gamma scheme had an RPS of 0.101 compared to the na\"{i}ve RPS of 0.579 and the RPS of the proposed method of 0.100 as presented in Section \ref{sec:Sim}.

\section{Bisquare Basis Comparison}
\label{Ap:BasisSensitivity}
As a sensitivity comparison, we considered bisquare basis functions with equally spaced knot points in the domain.
Due to the equal spacing imposed constraint, 6-knot locations were chosen, which is close to the number of EOFs but is a conservative comparison.
In this case, the model does not capture the evolving fire front as well, and the corresponding forecast predictions are worse when compared to the simulated EOF approach, with an RPS of 0.2502 compared to the method with simulated EOFs that had an RPS of 0.212.

\newpage
\setstretch{1} 
\bibliographystyle{apalike}
\bibliography{references_clean}

\begin{thebibliography}{}

\bibitem[Albert and Chib, 1993]{Albert1993BayesianData}
Albert, J.~H. and Chib, S. (1993).
\newblock {Bayesian Analysis of Binary and Polychotomous Response Data}.
\newblock {\em Journal of the American Statistical Association},
  88(422):669--679.

\bibitem[Alessandri et~al., 2020]{Alessandri2020ParameterMethods}
Alessandri, A., Bagnerini, P., Gaggero, M., and Mantelli, L. (2020).
\newblock {Parameter Estimation of Fire Propagation Models Using Level Set
  Methods}.
\newblock {\em Applied Mathematical Modelling}.

\bibitem[Andrews, 2018]{AndrewsTheExplanation}
Andrews, P.~L. (2018).
\newblock {The Rothermel Surface Fire Spread Model and Associated Developments:
  A Comprehensive Explanation}.
\newblock Technical report, U.S. Department of Agriculture.

\bibitem[Banks and Hooten, 2021]{Banks2021StatisticalModeling}
Banks, D.~L. and Hooten, M.~B. (2021).
\newblock {Statistical Challenges in Agent-Based Modeling}.
\newblock {\em American Statistician}, 75(3):235--242.

\bibitem[Cressie and Wikle, 2015]{cressie2015statistics}
Cressie, N. and Wikle, C.~K. (2015).
\newblock {\em Statistics for spatio-temporal data}.
\newblock John Wiley \& Sons.

\bibitem[Currie et~al., 2019]{Currie2019Pixel-levelSpread}
Currie, M., Speer, K., Hiers, J.~K., O’brien, J.~J., Goodrick, S., and
  Quaife, B. (2019).
\newblock {Pixel-level statistical analyses of prescribed fire spread}.
\newblock {\em Canadian Journal of Forest Research}, 49(1):18--26.

\bibitem[Dabrowski et~al., 2022]{dabrowski2022towards}
Dabrowski, J.~J., Huston, C., Hilton, J., Mangeon, S., and Kuhnert, P. (2022).
\newblock Towards data assimilation in level-set wildfire models using bayesian
  filtering.
\newblock {\em arXiv preprint arXiv:2206.08501}.

\bibitem[Epstein, 1969]{EpsteinRPS}
Epstein, E.~S. (1969).
\newblock A scoring system for probability forecasts of ranked categories.
\newblock {\em Journal of Applied Meteorology (1962-1982)}, 8(6):985--987.

\bibitem[Ferro et~al., 2008]{Ferro2008}
Ferro, C., Richardson, D., and Weigel, A. (2008).
\newblock On the effect of ensemble size on the discrete and continuous ranked
  probability scores.
\newblock {\em Meteorological Applications}, 15:19 -- 24.

\bibitem[Finney, 1998]{finney1998farsite}
Finney, M.~A. (1998).
\newblock {FARSITE}: Fire area simulator-model development and evaluation.
\newblock Technical report.

\bibitem[Gardner, 1970]{conway1970}
Gardner, M. (1970).
\newblock Mathematical games.
\newblock {\em Scientific American}, 223(4):120--123.

\bibitem[Hern{\'{a}}ndez~Encinas et~al.,
  2007]{HernandezEncinas2007ModellingAutomata}
Hern{\'{a}}ndez~Encinas, L., Hoya~White, S., Mart{\'{i}}n~del Rey, A., and
  Rodr{\'{i}}guez~S{\'{a}}nchez, G. (2007).
\newblock {Modelling forest fire spread using hexagonal cellular automata}.
\newblock {\em Applied Mathematical Modelling}, 31(6):1213--1227.

\bibitem[Hooten and Wikle, 2010]{hooten2010statistical}
Hooten, M.~B. and Wikle, C.~K. (2010).
\newblock Statistical agent-based models for discrete spatio-temporal systems.
\newblock {\em Journal of the American Statistical Association},
  105(489):236--248.

\bibitem[Hoover and Hanson, 2022]{hoover2022wildfire}
Hoover, K. and Hanson, L.~A. (2022).
\newblock Wildfire statistics.
\newblock Technical report, Congressional Research Service.

\bibitem[Johnston et~al., 2008]{Johnston2008EfficientGrid}
Johnston, P., Kelso, J., and Milne, G.~J. (2008).
\newblock {Efficient simulation of wildfire spread on an irregular grid}.
\newblock {\em International Journal of Wildland Fire}, 17(5):614--627.

\bibitem[Jolliffe and Stephenson, 2012]{jolliffe2012forecast}
Jolliffe, I.~T. and Stephenson, D.~B. (2012).
\newblock {\em Forecast verification: a practitioner's guide in atmospheric
  science}.
\newblock John Wiley \& Sons.

\bibitem[Karafyllidis and Thanailakis, 1997]{Karafyllidis1997AAutomata}
Karafyllidis, I. and Thanailakis, A. (1997).
\newblock {A model for predicting forest fire spreading using cellular
  automata}.
\newblock {\em Ecological Modelling}, 99(1):87--97.

\bibitem[Koo et~al., 2010]{Koo2010FirebrandsFires}
Koo, E., Pagni, P.~J., Weise, D.~R., and Woycheese, J.~P. (2010).
\newblock {Firebrands and spotting ignition in large-scale fires}.

\bibitem[Lautenberger, 2013]{Lautenberger2013WildlandCalibration}
Lautenberger, C. (2013).
\newblock {Wildland fire modeling with an Eulerian level set method and
  automated calibration}.
\newblock {\em Fire Safety Journal}, 62(PART C):289--298.

\bibitem[Liu et~al., 2018]{Liu2018SpreadSimulation}
Liu, Y., Liu, H., Zhou, Y., and Sun, C. (2018).
\newblock {Spread vector induced cellular automata model for real-time crown
  fire behavior simulation}.
\newblock {\em Environmental Modelling and Software}, 108:14--39.

\bibitem[Mallet et~al., 2009]{Mallet2009ModelingMethods}
Mallet, V., Keyes, D.~E., and Fendell, F.~E. (2009).
\newblock {Modeling wildland fire propagation with level set methods}.
\newblock {\em Computers and Mathematics with Applications}, 57(7):1089--1101.

\bibitem[Mason, 2004]{MasonRPSS}
Mason, S.~J. (2004).
\newblock On using “climatology” as a reference strategy in the brier and
  ranked probability skill scores.
\newblock {\em Monthly Weather Review}, 132(7):1891 -- 1895.

\bibitem[Mu{\~{n}}oz-Esparza et~al., 2018]{Munoz-Esparza2018AnMethod}
Mu{\~{n}}oz-Esparza, D., Kosovi{\'{c}}, B., Jim{\'{e}}nez, P.~A., and Coen,
  J.~L. (2018).
\newblock {An Accurate Fire-Spread Algorithm in the Weather Research and
  Forecasting Model Using the Level-Set Method}.
\newblock {\em Journal of Advances in Modeling Earth Systems}, 10(4):908--926.

\bibitem[Ottmar et~al., 2015]{RXcadreExperiment}
Ottmar, R., Hiers, J., Butler, B., Clements, C., Dickinson, M., Hudak, A.,
  O'Brien, J., Potter, B., Rowell, E., Strand, T., and Zajkowski, T. (2015).
\newblock Measurements, datasets and preliminary results from the rxcadre
  project ? 2008, 2011 and 2012.
\newblock {\em International Journal of Wildland Fire}, 25.

\bibitem[Polson et~al., 2013]{polson2013bayesian}
Polson, N.~G., Scott, J.~G., and Windle, J. (2013).
\newblock Bayesian inference for logistic models using p{\'o}lya--gamma latent
  variables.
\newblock {\em Journal of the American statistical Association},
  108(504):1339--1349.

\bibitem[Rothermel, 1972]{RothermelRAFuels.}
Rothermel, R.~C. (1972).
\newblock {A mathematical model for predicting fire spread in wildland fuels.}

\bibitem[Schaefer, 1990]{Schaefer1990}
Schaefer, J.~T. (1990).
\newblock The critical success index as an indicator of warning skill.
\newblock {\em Weather and Forecasting}, 5(4):570 -- 575.

\bibitem[Schliep and Hoeting, 2015]{SCHLIEP20151}
Schliep, E.~M. and Hoeting, J.~A. (2015).
\newblock Data augmentation and parameter expansion for independent or
  spatially correlated ordinal data.
\newblock {\em Computational Statistics \& Data Analysis}, 90:1--14.

\bibitem[Simmonds et~al., 2020]{Simmonds2020TheReview}
Simmonds, J., G{\'{o}}mez, J.~A., and Ledezma, A. (2020).
\newblock {The role of agent-based modeling and multi-agent systems in
  flood-based hydrological problems: A brief review}.
\newblock {\em Journal of Water and Climate Change}, 11(4):1580--1602.

\bibitem[Stephenson, 2000]{Stephenson2000}
Stephenson, D.~B. (2000).
\newblock Use of the “odds ratio” for diagnosing forecast skill.
\newblock {\em Weather and Forecasting}, 15(2):221 -- 232.

\bibitem[Sullivan, 2007]{Sullivan2007AModels}
Sullivan, A.~L. (2007).
\newblock {A review of wildland fire spread modelling, 1990-present 2:
  Empirical and quasi-empirical models}.
\newblock {\em International Journal of Wildland Fire}.

\bibitem[von Neumann, 1966]{JohnVonNeumannTheoryAutomata}
von Neumann, J. (1966).
\newblock Theory of self-reproducing automata.
\newblock {\em Edited by Arthur W. Burks}.

\bibitem[Wikle and Hooten, 2015]{wikle2015hierarchical}
Wikle, C.~K. and Hooten, M.~B. (2015).
\newblock Hierarchical agent-based spatio-temporal dynamic models for
  discrete-valued data.
\newblock {\em Handbook of Discrete-Valued Time Series}, pages 349--366.

\bibitem[Wikle et~al., 2019]{wikle2019spatio}
Wikle, C.~K., Zammit-Mangion, A., and Cressie, N. (2019).
\newblock {\em Spatio-temporal Statistics with R}.
\newblock Chapman and Hall/CRC.

\bibitem[Yoo and Wikle, 2022]{YooLevelSet}
Yoo, M. and Wikle, C.~K. (2022).
\newblock A bayesian spatio-temporal level set dynamic model and application to
  fire front propagation.

\bibitem[Zhang and Wang, 2021]{Zhang2021AModels}
Zhang, B. and Wang, H. (2021).
\newblock {A new type of dual-scale neighborhood based on vectorization for
  cellular automata models}.
\newblock {\em GIScience and Remote Sensing}, 58(3):386--404.

\end{thebibliography}

\end{document}